# Measuring Night-Sky Brightness with a Wide-Field CCD Camera

Dan M. Duriscoe

Death Valley National Park, Death Valley, CA; dan_duriscoe@nps.gov

Christian B. Luginbuhl

US Naval Observatory Flagstaff Station, Flagstaff, AZ; cbl@nofs.navy.mil

AND

Chadwick A. Moore

Bryce Canyon National Park, Bryce Canyon, UT; chad_moore@nps.gov

**ABSTRACT.** We describe a system for rapidly measuring the brightness of the night sky using a mosaic of CCD images obtained with a low-cost automated system. The portable system produces millions of independent photometric measurements covering the entire sky, enabling the detailed characterization of natural sky conditions and light domes produced by cities. The measurements are calibrated using images of standard stars contained within the raw data, producing results closely tracking the Johnson *V* astronomical standard. The National Park Service has collected hundreds of data sets at numerous parks since 2001 and is using these data for the protection and monitoring of the night-sky visual resource. This system also allows comprehensive characterization of sky conditions at astronomical observatories. We explore photometric issues raised by the broadband measurement of the complex and variable night-sky spectrum, and potential indices of night-sky quality.

## 1. INTRODUCTION

The National Park Service's mission of preservation of national park resources and values requires accurate information on past, present, and future conditions. The protection of night-sky quality, or "lightscape management," has been mandated in the agency's management policies (NPS 2006). While the Moon produces by far the greatest amount of natural illumination, many nocturnal species have evolved to take advantage of moonless nights between dusk and dawn (sometimes referred to by astronomers as "dark time"). In addition, a significant component of the natural visual scene in many national parks is the appearance of the night sky on a dark night, and the National Park Service is charged with the duty of conserving the scenery and the wildlife within parks (16 U.S.C., chaps. 1, 2, 3, and 4). Monitoring sky brightness and the sources contributing to it can best be accomplished by measurements of the entire sky; that is, from the zenith to the horizon in all azimuthal directions. This is because light from the entire sky contributes to illumination of the land, and visitors to natural areas such as these will easily perceive light domes from distant cities as degradation of the natural night sky, even if they affect only a small portion of the sky that is of little interest to astronomers.

All-sky images obtained with fish-eye lenses have been employed by astronomical observatories for cloud monitoring and qualitative or semiquantitative representations of sky quality (such as the TASCA camera at Cerro Tololo Inter-American Observatory [Schwarz et al. 2003] and the network of continuous night-sky cameras known as CONCAM [Pereira & Nemiroff 1998]), but such images are difficult to calibrate to known standards of sky brightness measurement. Calibrated broadband sky brightness measurements have traditionally been made at major observatories at specific points in the sky, either with a single-channel photoelectric photometer or a narrow-field (i.e., astronomical telescope) CCD detector (e.g., see Walker 1970, 1973; Pilachowski et al. 1989; Mattila et al. 1996; Nawar 1998; Lockwood et al. 1990; Patat 2003). The approach described in this paper represents an innovative use of wide-field CCD images for panoramic sky brightness photometry, with an image scale that is sufficient to perform the accurate stellar photometry required for calibration to known standards. A similar system is under development by Cinzano & Falchi (2003). Our system provides all-sky coverage with the ability to combine data from multiple images into a single representation similar to that obtained with a fish-eye lens, while maintaining sufficient precision and accuracy to assess quantitative effects and long-term trends of light pollution. This product is achieved entirely with off-the-shelf, commercially available hardware at a cost of under $15,000 US.

Previous researchers of this subject have not discussed the complexity of measuring a night-sky spectrum characterized by emission lines using a broadband system calibrated with continuous stellar spectra. In addition, the relationship of such measurements to the visual perception of night-sky brightness by the dark-adapted human eye has received little attention. We provide a discussion that is pertinent to our photometric system and the $V$ band of the Johnson $UVB$ system.

TABLE 1
List of Equipment for Field Portable Sky Brightness Measurements

| Item | Weight (kg) | Brand Name and Model |
|---|---|---|
| Tripod ......................... | 4 | Celestron NexStar |
| Robotic telescope mount ...... | 6 | Celestron NexStar 5i |
| CCD camera ................... | 2–4 | Apogee AP260EP |
| | | Finger Lakes Instruments MaxCam CM9 |
| | | SBIG STL 1001E |
| | | Finger Lakes Instruments IMG 1001E |
| Camera lens ................... | 0.5–1 | Nikon 35 mm, f/1.4 |
| | | Nikon 50 mm, f/1.8 |
| Camera filter ................... | 0.1 | Custom Scientific Bessell $V$ |
| Notebook computer ............ | 3 | Dell Inspiron |
| | | IBM ThinkPad |
| Cables (three) ................. | 1 | USB, serial, 12 V DC power |
| Power supply .................. | 2–4 | Regulated DC-to-DC converter powered by a 15 V, 12 Ah Ultralife Lithium-ion battery (varies with camera; SBIG STL camera does not require regulated DC supply) |
| GPS receiver .................. | 0.3 | Garmin Etrex |
| Hand-held weather meter ...... | 0.1 | Kestrel 3500 |
| Total weight: | 19–24 | |

An important tool in managing light pollution is the use of models of the propagation of light in the atmosphere from sources on the ground. This work will contribute significantly to the validation of such models by providing numerous data points throughout the hemisphere of the sky (see Cinzano & Elvidge 2004). Its field portability also allows for the collection of data from many locations in the vicinity of a light source.

## 2. THE DATA ACQUISITION SYSTEM AND DATA PROCESSING

### 2.1. Data Quality Objectives

We set the following objectives for data quality for this monitoring effort:

1. Sky background brightness measurements must cover the entire sky with a resolution of no less than 2°.

2. Under the darkest of natural conditions, measurements must have a precision and accuracy of ±10% or better.

3. Measurements must be related to a known astronomical system to allow calibration of the data using stellar sources, and must closely approximate the human visual perception of brightness.

4. The extinction coefficient of the atmosphere must be determined for each data set.

5. The time interval between the beginning and ending of a single all-sky data set must not exceed 40 minutes.

The reasons for setting these objectives are as follows:

1. Higher resolution observations allow the size, shape, and total brightness of light domes to be characterized accurately.

2. Experiments with the equipment described herein revealed that to achieve precision greater than ±10% under the darkest of conditions (greater precision is easily attained under moderately bright and bright skies), modifications to the method would not allow several of the other objectives described to be met, particularly the completion of an all-sky data set within 40 minutes.

3. The Johnson $V$ band is an acceptable estimator of human visual perception of night-sky brightness, and the Johnson $UVB$ photometric system is well established, with published stellar magnitudes in many star catalogs.

4. Sky brightness resulting from distant ground sources is strongly dependent on atmospheric extinction, and any analysis or model validation will require a measurement of the extinction coefficient as a covariable.

5. The need for as brief an interval as possible between the beginning and ending of an all-sky data set is predicated on the fact that the brightness of portions of the sky, illuminated from both natural and artificial sources, can change rapidly—40 minutes was considered an acceptable compromise between a true "snapshot" and the limitations of the equipment.

### 2.2. Hardware

A suite of equipment required for field data collection was selected based on the following criteria: relatively low cost, transportable by one or two persons via foot, off-the-shelf availability, and durability. The equipment sets evaluated in this study are listed in Table 1.

#### 2.2.1. Camera

The heart of this system is the camera, including the CCD detector, associated electronics, and thermoelectric cooler. Cameras used in the development of these methods employed

TABLE 2
PROPERTIES OF THE CCD DETECTORS

| Camera | Detector | Array Size (pixels) | Pixel Size ($\mu$m) | Full-Well Depth ($e^-$) | Gain ($e^-$ ADU$^{-1}$) | Zero Point (mag) | Color Term (mag) | Plate Scale (arcsec pixel$^{-1}$) |
|---|---|---|---|---|---|---|---|---|
| Apogee AP260EP* ........... | KAF 261E | 256 × 256 | 40 × 40 | 800,000 | 1.8 | 14.00 ± 0.03 | −0.035 ± 0.005 | 231.5 |
| FLI MaxCam CM9 #1* ...... | KAF 261E | 256 × 256 | 40 × 40 | 800,000 | 3.85 | 13.40 ± 0.03 | −0.035 ± 0.005 | 231.5 |
| FLI MaxCam CM9 #2* ...... | KAF 261E | 256 × 256 | 40 × 40 | 800,000 | 1.47 | 14.40 ± 0.03 | −0.035 ± 0.005 | 231.5 |
| SBIG STL ................... | KAF 1001E | 1024 × 1024 | 24 × 24 | 200,000 | 1.96 | 14.79 ± 0.03 | −0.045 ± 0.006 | 93.6 |
| FLI IMG ..................... | KAF 1001E | 1024 × 1024 | 24 × 24 | 200,000 | 2.35 | 14.69 ± 0.03 | −0.045 ± 0.006 | 93.6 |

NOTE.—Asterisks denote cameras whose values reflect 2 × 2 on-chip binning.

two different detectors (hereafter referred to as "small format" and "large format," respectively): the 512 × 512 pixel Kodak KAF 261E and the 1024 × 1024 pixel Kodak KAF 1001E; both are front-illuminated. These detectors were selected because of their relatively low cost (especially the small-format), extended blue sensitivity, large full-well capacity (resulting in a large dynamic range and allowing accurate photometric measurements of both the sky background and stars on the same image), and large pixel size (resulting in high sensitivity per unit area of sky and allowing relatively short integration times). We used 2 × 2 binning with the small-format systems to increase the effective sensitivity; the large-format system performed adequately unbinned. The sensors' key properties are listed in Table 2.

The small-format camera was purchased for initial development of the system because it met the cost and availability criteria in 2000. Steady technological improvements and cost reductions in high-end consumer astronomical CCD cameras allowed the large-format system to become commercially affordable in 2004, and two cameras using this detector were purchased, with the intent of evaluating their potential advantages. In total, five separate cameras from three different man-ufacturers (Apogee Instruments, Santa Barbara Instrument Group [SBIG], and Finger Lakes Instrumentation [FLI]) were purchased and tested. Each camera has its own particular characteristics, requiring customized data-processing methods (Table 2).

While professional-grade CCD cameras typically exhibit a very linear response to photon flux, deviations from linearity in these semiprofessional or amateur-grade cameras often occur at the highest and lowest ends of their dynamic range. Nonlinearity at low light levels was of particular concern here, because measurements of the sky background brightness at sites that are remote from light pollution typically produce a very low signal with optical systems and integration times that may be practically employed for this project. Deviations from a linear response were measured under laboratory conditions through the use of a voltage-regulated LED light source and varying exposure times. These tests revealed that the FLI large-format camera exhibited excellent linearity from 70 to 62,000 ADUs pixel$^{-1}$, the SBIG large-format camera showed a significant decrease in sensitivity below 4000 ADUs pixel$^{-1}$, while all the small-format cameras had a drop-off in sensitivity below 10,000 ADUs pixel$^{-1}$ (Fig. 1). At the upper end of their dynamic range, each detector also displayed an expected falloff in response due to pixel saturation. The loss of low-end sensitivity is apparently due to diffusely distributed electron traps with time constants above the pixel read time of the amplifier (F. J. Harris 2005, private communication). The effect was more severe at lower temperatures and was found to be alleviated somewhat by 2 × 2 on-chip binning in the small-format cameras. Each camera produced a slightly different linearity curve, even those that used the same detector, presumably because of variations in the manufacturing process and the electronics used in cameras of different manufacturers. All cameras output 16 bit integer ADUs, so their maximum range is 0–65,535.

Nonlinearity in response to lower photon flux would significantly compromise the ability of these cameras to accurately measure sky background brightness. Therefore, empirically derived linearizing functions were applied pixel-by-pixel as part of the image calibration process to adjust low-ADU pixels. The nonlinearity of response to low photon flux was observed to be very reproducible for a given detector temperature, and adjustments made to low measured ADU values by these func-tions introduce very little error. Longer exposures could alleviate the need for using these functions by increasing the low light signal. However, it was necessary to limit the exposure time, because both the sky background brightness and standard stars are measured on the same image, and thus pixels containing the brighter standard stars used for extinction coefficient calculation would become saturated when exposures longer than 15 s were attempted.

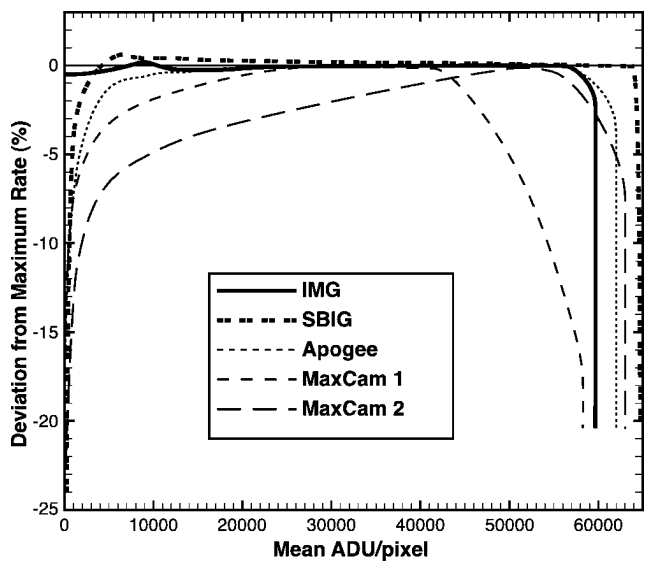

FIG. 1.—Deviations from linear response for cameras employed in this study.

A thermoelectric cooler allows for higher quality images by reducing thermal noise from the CCD detector. However, at a remote location, there is a trade-off between increased image quality and the amount of battery power required to achieve colder detector temperatures. Therefore, nominal operating temperatures of $-10°C$ for the small-format and $-20°C$ for the large-format systems were selected. Under nearly all conditions, these temperatures are easily reached by the camera's coolers. Occasionally, during winter observations at high-elevation locations, the cameras had to be set to a colder temperature because the ambient temperature was below the nominal operating temperature. Power consumption of the cameras ranges from 1 to 4 A at 12 V DC, depending on the camera and the amount of cooling required. The 12 Ah lithium ion battery will typically allow for 4–6 hr of continuous image acquisition.

### 2.2.2. Optics

The objectives of the method require the use of the widest angle lens possible while still maintaining the ability to perform stellar photometry with reasonable precision. We found that a lens of f/2 or faster appears to be necessary to produce an adequate signal-to-noise ratio on sky brightness measurements under the darkest natural skies with ≤15 s exposures. Typical exposure times employed are 10 s for the small-format cameras and 12 s for the large-format cameras. Lenses of 35 and 50 mm focal length were determined to be the shortest focal length possible for the small- and large-format systems, respectively, both operating at f/2. Standard Nikon camera lenses were selected and adapted for use with the CCD cameras. These lenses produce pixel scales of 231.5″ and 93.9″ for the small-(binned 2 × 2) and large-format cameras, respectively. With these short focal lengths, stellar images were small enough that the location of the image relative to gate structures on the front-illuminated CCDs had large effects on the photometric precision. This problem was overcome by introducing a slight defocus of the camera, spreading the stellar images over a large enough area that gate-structure effects averaged out enough to reduce the error that was introduced to about 0.01 mag. In addition, the images are made with the camera fixed (sidereal tracking off) once it is properly pointed, allowing stars away from the pole to blur slightly as they move across the sky. With a 12 s exposure, the trailing of stellar images is less than 2 pixels with the large-format camera for a star at the celestial equator.

### 2.2.3. Filter

When choosing the detector and filter system, compromises were necessary to (1) allow photometric calibration using available astronomical standards, (2) maintain a clear relation to previous sky brightness measurements published in the astronomical literature, and (3) approximate complex human visual response under the wide range of night sky conditions from pristine national parks to locations with considerable artificial sky glow caused by a varying mixture of artificial light sources.

The choice was made to closely approximate the Johnson $V$ (henceforth referred to as $V_J$) response (Johnson & Morgan 1953) by using a Bessell $V$ filter. To maximize uniformity for each of the five camera systems, five 50 mm filters cut from the same batch of glass were obtained from Custom Scientific, Inc. The filter includes an infrared blocker. Section 2.3 briefly discusses the photometric characteristics of the system, which are more thoroughly examined in § 4.1.

### 2.2.4. Telescope Mount

The camera must be pointed accurately in altitude and azimuth to the target point of each image comprising an all-sky data set. A small, lightweight, robotic telescope mount allowed the pointing process to be completely automated. The Celestron NexStar 5i was chosen primarily for its ability to quickly slew from one position to the next, accomplishing the acquisition of an all-sky data set in less than 35 minutes. Its pointing accuracy was found by testing to be within 0.5°. This mount can be computer controlled, with pointing in both equatorial and altitude-azimuth coordinates available. An aluminum adapter plate to which the camera is fastened (Fig. 2) replaces the optical tube assembly for which the mount is designed. While this mount is capable of sidereal tracking, all images were exposed with tracking off.

### 2.2.5. Computer Control and Image Retrieval

A notebook PC computer provides the means for pointing the telescope, controlling the image acquisition program, and retrieving and organizing images from each data set at a field location. We selected off-the-shelf notebook computers from Dell and IBM running the Windows XP operating system. Computers with two built-in batteries allow for sessions of up to 6 hr; longer observing runs are possible with an external battery and a 12 V DC power adapter. Standard lithium ion computer batteries work well in all conditions; lithium polymer batteries, which are now provided with some notebook computers as secondary batteries, should be avoided in cold weather conditions unless the computer can be kept warm. The XP operating system allows the computer's monitor to be turned off during operation, significantly extending battery life. Cables from the computer to the camera and telescope mount allow for communication between the devices; these cables must be of sufficient length and flexibility to allow for the mount's 360°

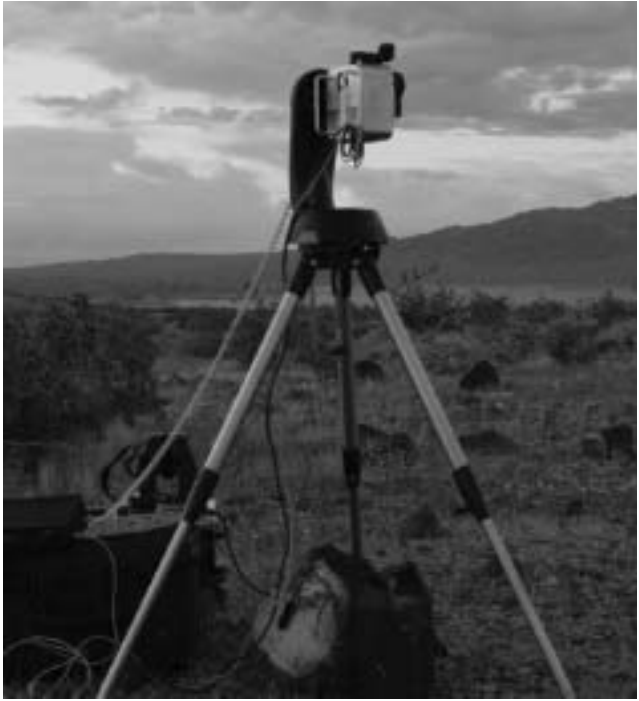

Fig. 2.—Field setup of one of the large-format cameras.

rotation in azimuth without binding, a nontrivial problem under cold conditions.

## 2.3. System Spectral Response

A detailed analysis of our system's spectral characteristics and its response to natural and artificial light sources in the night sky is presented in § 4.1. Here we provide a brief summary. Henceforth, we refer to our system's response as the NPS (National Park Service) or $V_{NPS}$ system, and to magnitudes made on the system as $V_{NPS}$.

The response of the NPS system is compared to human photopic (light adapted, 2° foveal field of view; CIE 1926) and scotopic (completely dark adapted, 10° foveal field of view; CIE 1964) visual response in Figure 3, while the effective wavelengths and full widths to both 50% and 10% responses are listed in Table 3. (We present the filtered KAF 261E response as representative; differences between the KAF 261E and KAF 1001E detectors are discussed in § 4.1) The NPS system exhibits an effective wavelength that is intermediate between the two visual response functions, with a comparable bandwidth. Although it might be argued that a more appropriate response for our system would have been one more closely approximating either of the two visual responses, it is important to recognize that we are measuring sky conditions spanning a range of brightness in which both scotopic and photopic visual responses are to be expected, from naturally dark night skies to heavily polluted sites located near large cities, or for observation points close to the horizon in the direction of large cities. Under all but the darkest (natural) conditions, eye response will be intermediate between photopic and scotopic limits, in the so-called mesopic range. There is no single mesopic response function, since the spectral response of the eye under intermediate lighting levels is dependent on more variables than simply the lighting level. Most relevant here is the spectral content of the light, which naturally can be expected to vary tremendously when comparing natural sources to artificial sources, and when recognizing the large variety of artificial sources in common use. Thus, an intermediate response such as that chosen here for the NPS system seems a useful compromise for characterizing the visual appearance of the sky under a wide range of conditions.

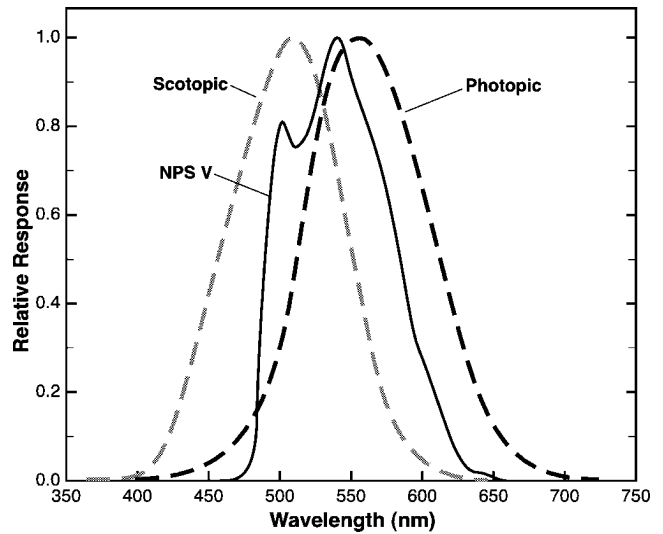

Fig. 3.—Relative spectral response of the NPS $V$ band (*solid line*) and human photopic (*black dashed line*) and scotopic (*gray dashed line*) vision.

Figure 4 shows the relative response of the $V_{NPS}$ and $V_J$ systems, normalized for equal area under the curves. This figure and the values in Table 3 show that the NPS system response closely approximates the standard Johnson $V$ response. The slight difference in effective wavelength will require the use of a correction factor (an instrumental color term) when comparing stellar measurements made on our system to $V_J$, a prac-

### TABLE 3
### System Effective Wavelengths and Full Widths

| System | $\lambda_{eff}$ (nm) | Full Width (nm) | |
| --- | --- | --- | --- |
| | | 50% | 10% |
| Scotopic | 502 | 95 | 162 |
| $V_{NPS}$ | 541 | 98 | 128 |
| $V_J$ | 551 | 92 | 153 |
| Photopic | 560 | 100 | 179 |

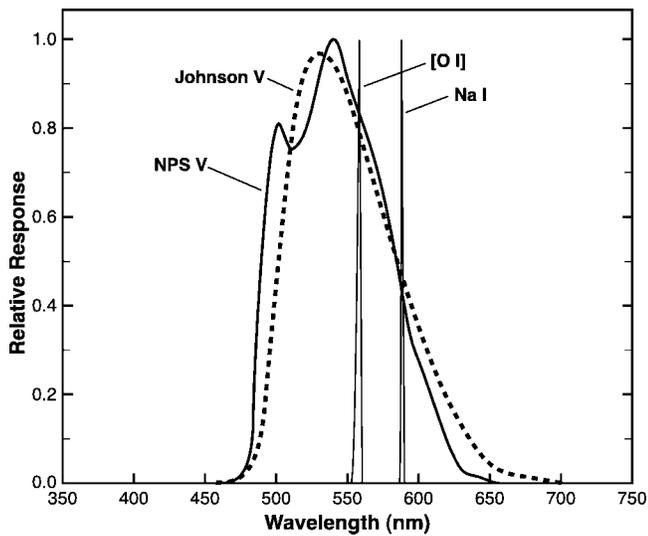

Fig. 4.—Relative spectral response of the NPS (*solid line*) and $V_J$ (*short dashed line*) bands. Two emission lines important to the night-sky spectrum (Na I at 589 nm and [O I] at 557.7 nm) are indicated for reference.

tice that is ubiquitous in astronomical photometric measurements when transforming measurements to a standard system defined with different equipment. This small correction, while useful for characterizing the relationship of our system to the standard $V_J$ system, is used only to remove color effects when determining instrumental zero-point and atmospheric extinction values, and not applied to measurements of sky brightness, since calibration factors based on stellar measurements cannot be used to correct measurements of sources such as sky glow with substantially different spectral energy distributions than those exhibited by stars. Nonetheless, since the $V_{NPS}$ and $V_J$ spectral responses are quite similar, sky brightness measurements on the $V_{NPS}$ system will be quite close to those made by a hypothetical system exactly duplicating the $V_J$ response (see § 4.1).

The two narrow emission features depicted in Figure 4 are important components of the night-sky spectrum. The [O I] line at 557.7 nm results from the natural airglow, while the Na I emission at 589 nm primarily originates from low-pressure sodium (LPS) vapor lights. It can be seen from the graph that the airglow component should be measured with an intensity very close to a detector with a true $V_J$ response, while LPS and to a similar extent high-pressure sodium (HPS) sources will be slightly underestimated in $V_{NPS}$ relative to $V_J$.

### 2.4. Field Data Acquisition

Observing sites are selected with as much unobstructed view of the horizon in all directions as is possible, and/or at locations that are well known or important to the experience of visitors to national park areas (such as Glacier Point in Yosemite Na-tional Park). The equipment is carried by hand or backpack to each observing site, which is generally located within a few hundred meters of a road but may be several kilometers from the nearest road and in designated wilderness.

Telescope setup involves the accurate alignment of the altaz mount to true north and to level, and initialization of the NexStar software to the site's exact location, date, and time. This is accomplished by imaging Polaris to determine a zero point for azimuth, and by centering a bubble level placed on top of the camera case while rotating the instrument in azimuth to ensure that the azimuthal axis is vertical. Accurate Universal Time is obtained from a hand-held GPS receiver, as is the precise latitude and longitude of the location. The computer's clock is also synchronized to the GPS before each observing session begins. Air temperature, relative humidity, and wind speed are observed with a hand-held weather meter. These observations, along with camera type, the exposure time in seconds, number of data sets desired, and the interval between data sets in minutes are typed into the notebook computer by the observer just before data collection begins.

Image acquisition, including bias, dark, and sky frames, proceeds automatically, controlled by a Visual Basic script and the appropriate software (see § 2.5), which coordinates the operations of telescope mount and camera. The all-sky imaging program for each data set obtains five initial bias and dark frames, a single image of the zenith at the start of the data set, systematic sequential images that most efficiently cover the entire sky (104 images for the small-format system, 45 images for the large-format), bias frames interspersed between the sky images to check for bias drift, and a second zenith image at the end of the data set. Each image includes at least 1° of overlap with adjacent images so that no part of the sky is omitted. The procedure takes about 35 minutes to complete with the small-format system, and about 22 minutes with the large-format. All of the attributes that were typed in by the observer are recorded in the header of each image, along with target right ascension, declination, altitude and azimuth, and the date and time the exposure begins. Image files are saved in a 16 bit integer FITS format.

Up to eight sequential data sets are obtained automatically, with a user-defined interval between each. A 15 V, 12 Ah lithium ion battery is sufficient to allow the completion of at least five sets at 1 hr intervals. In this manner, changes in sky brightness and extinction properties can be monitored throughout the night.

### 2.5. Software and Processing

#### 2.5.1. Image Acquisition

Commercial software for the Windows XP operating system was selected that allowed fully automated data acquisition. MaxIm DL, version 4.06, from Diffraction Limited software, was selected as the primary camera control and image-pre-processing program. ACP Observatory Control Software, by DC-3 Dreams software, was selected to coordinate camera and telescope control. Both programs provide a large suite of scriptable commands and functions that allowed the customization required for full automation of this observing program. Scripts were written in Visual Basic to systematically step through the process of pointing the telescope, acquiring images, and organizing the retrieved image files on the computer's hard disk for automated processing later on.

### 2.5.2. Image and Data Processing

The tools required for automated image processing (including bias and thermal signal subtraction, flat-field scaling, stellar photometry, and sky background measurements) were available in MaxIm DL. Custom Visual Basic scripts step through the processing procedure and output measurements to Microsoft Excel spreadsheets in the form of data tables. Further calculations are performed in the spreadsheets. Certain astronomical calculations required utilities provided by ACP Observatory Control Software. Individual images were solved astrometrically with the astrometric engine PinPoint 4, by DC-3 Dreams software, providing plate solutions and the ability to locate and extract photometric measurements of standard stars for calibration purposes. Graphic display of the mosaicked all-sky data was accomplished by creating interpolated contour maps in ArcView Spatial Analyst, by Environmental Systems Research Institute.

Full processing involves the following steps: (1) frame calibrations (linearization, bias/thermal frame subtraction, and flat-field correction), (2) astrometric image solution, (3) standard-star measurement, (4) standard-star processing, (5) sky brightness measurement, (6) sky brightness processing, (7) creation of contour maps of sky brightness, and (8) calculation of indices of sky brightness. Each step is described in detail below.

### 2.5.3. Frame Calibrations

Standard image calibration procedures, including the subtraction of bias and thermal signal and division by a master flat, were performed on all-sky images. In addition, a correction for nonlinear response at the low end of the dynamic range of each camera exhibiting this effect (described in § 2.2) was applied to both thermal and photon signal. Observed bias drift was also corrected by referencing to the interspersed bias frames.

The relatively wide field of view of the large-format systems presented challenges for obtaining accurate flat-field frames. The master flat for these cameras was formed by imaging a diffusing screen placed 1 cm from the front of the lens, which was in turn illuminated by light from the daytime sky. The camera was set to its nominal operating temperature, and the Bessel V filter and the same f-stop that is used for field data collection (f/2) were employed. Sufficient illumination was provided to achieve an average signal of about 50% of the max-imum pixel value, well within the linear range of each camera's response curve. An integration time of at least 2 s was used to minimize shutter effects on the flat. Even with full control over these procedures, variations in replicate flat images were observed for the large-format camera. The physical size of its detector, combined with the f/2 focal ratio and wide field of view (26° edge to edge), results in a large amount of vignetting from center to corner (nearly 50%). Correcting this variable to within 1% or better represented a considerable challenge, requiring a good deal of trial and error in positioning the diffusing screen. Ultimately, an average of many individual flat frames was constructed. This master flat was tested by mosaicking images from a night-sky data set and examining the seams for variations in brightness at all four edges of the frame. In this manner, it was determined that the flat succeeded in correcting the vignetting to within 1%.

### 2.5.4. Astrometric Image Solution

An attempt to solve each image astrometrically was made in batch mode with the PinPoint 4 Astrometric Engine. A solution could not be obtained for all images, especially those with directly visible city lights, trees, or other obstructions. Lights below the horizon were masked by blocking those parts of the image that lie below 5° altitude. In addition, occasionally astrometric solutions could not be obtained for images that included bright light domes, presumably because of the severe sky background gradient across the frame. Once solved, solution parameters were written to the fits header of each image and used in future processing steps to find standard stars.

### 2.5.5. Standard-Star Measurement

The atmospheric extinction coefficient can vary significantly from night to night, or even during a night. Although its value is not directly relevant for determining sky brightness measurements (i.e., sky brightness measurements are not "brightened" by removing the effects of extinction), it is an important factor in the propagation of both natural and artificial light through the atmosphere, and therefore also the sky brightness characteristics of an unpolluted sky, as well as the effect of light pollution sources. An accurate determination of its value for each data set is an important part of this method. In addition, the instrumental zero point for each camera/temperature combination is determined by the $y$-intercept of the extinction curve after color effects have been removed (see below). An accurate determination of these values is therefore critical to proper interpretation of sky brightness measurements.

Standard stars were first selected on the basis of their brightness. As mentioned above, the brightness of both sky background and stars must be measured accurately on the same image. Images of stars brighter than $V_i = 3.8$ were observed to nearly always yield at least 1 pixel with ADU values greater than 40,000 with either format camera, thus defining the upper brightness limit. The lower brightness limit for standard stars

2007 PASP, **119**:192–213

was set at $V_J = 6.0$ to provide a high signal-to-noise ratio. Therefore, a list of 391 standard stars with $3.8 \leq V_J \leq 6.0$ that are not known variable stars was selected from the *Hipparcos* catalog (ESA 1997). Close companions listed in the Washington Double Star Catalog (Mason et al. 2001), and stars within the boundaries of the bright sections of the Milky Way, were excluded. *Hipparcos* photometric values, including the $V_J$ magnitude and $(B_J - V_J)$ color index, were used to determine a fixed instrumental zero point and color term for each camera system (see below), in addition to the atmospheric extinction coefficient for each data set.

A Visual Basic procedure using commands and functions in the MaxIm DL and PinPoint software packages is used to find all standard stars located above 5 air masses (approximately 78° zenith distance) on each astrometrically solved image and measure their apparent brightness with synthetic aperture photometry. The corners of the large-format images are excluded because of excessive distortion of the stellar images in these regions. The measurement methods follow those outlined for CCD images by Howell (2000). Stellar flux is summed within a circular aperture centered on each star; the aperture radius is set to 4 pixels (15.4′) for the small-format system, and 6 pixels (9.6′) for the large-format system; typical FWHM values average 1.4 and 2.0 pixels, respectively. The measured star intensities are converted to instrumental magnitude on a spreadsheet. In this manner, between 70 and 130 stars are automatically measured per data set.

### 2.5.6. Standard-Star Processing

The same procedure described above calculates the zenith angle of each star from the latitude, longitude, date, and time of each image (retrieved from the FITS header), and the output spreadsheet converts it to air mass. A table that includes $V - v$ (published magnitude minus instrumental magnitude), air mass, and $(B - V)$ color index is automatically entered into an Excel spreadsheet. The extinction coefficient, color term, and zero point are determined via multiple linear regression using the equation

$$V - v = Q - k'_v X + \epsilon(B - V),$$

where $V$ is the $V_J$ magnitude of the standard star from the *Hipparcos* catalog, $v$ is the instrumental magnitude of the star, $Q$ is the instrumental zero point, $k'$ is the extinction coefficient, $\epsilon$ is the instrumental color transformation coefficient, $X$ is the air mass, and $(B - V)$ is the color index from the *Hipparcos* catalog.

The first-order color term was observed to be constant within the uncertainties for a given detector/temperature combination for both the large- and small-format cameras. This is not unexpected, since the spectral response characteristics of the system reflected in this term are expected to be stable, provided changes in the throughput of the system (primarily affecting the instrumental zero point $Q$) exhibit no color dependence. We determined a color transformation coefficient for each camera, using an average of the values obtained on at least five nights (see Table 2), following the methods described by Sung & Bessell (2000). For all final reductions, the color transformation coefficient was held fixed. Color corrections are not applied to measurements of sky brightness.

The instrumental zero point $Q$ for each camera is also generally stable but was occasionally observed to vary slightly. This term is sensitive to the total throughput of the system and therefore is affected by the cleanliness of the filter and camera optics and small variations in the aperture of the camera lens. Therefore, the zero point is checked each night with an unconstrained least-squares regression. If the unconstrained reduction indicates a significantly different zero point from the long-term average, and if this difference appears to be consistent for all of the data sets obtained during a particular night, the least-squares value is used instead of the average. The optics are then checked for cleanliness and proper function of the iris diaphragm.

### 2.5.7. Sky Background Brightness Extraction and Measurement

The brightness of the sky background on CCD images is determined by taking the median value of a circular region approximately 1° in diameter (16 pixels for the small-format = 1.03°, and 40 pixels for the large-format = 1.04°). The median is the only measurement statistic available for this purpose using MaxIm DL software. The mode statistic, lowest quartile, or "mean of median half" may describe the sky background more accurately, but the median statistic typically produces nearly the same value, unless a very bright star or planet with a large image area or halo is included in the area measured (Berry & Burnell 2000). The larger the area measured, the more "sky" is included and the less important stellar images become. An examination of our data reveals that virtually all star images are small enough to be filtered out, but the very brightest objects, such as Sirius, Vega, Jupiter, and Venus, often produce images that are large enough to bias the median above the sky background. Therefore, these objects may produce single points with large errors in sky brightness measurement. Extended "natural" bright areas, such as Milky Way star clouds, zodiacal light, and bright nebulae, are measured as part of the sky background brightness by our method, and no attempt is made to subtract them at this time. This method extracts a single value corresponding to a particular azimuth and altitude, which, while having no true replicate measures in our system, represents data from hundreds of pixels with an overall high signal-to-noise ratio. For a simplified analysis and a manageable database, we selected sample points evenly distributed every 2°. In this manner, 20% of the total image area is sampled. These measurements are made systematically centered on predetermined pixel addresses corresponding to the appropriate altitude and azimuth coordinates, using a Visual Basic script and tools available in MaxIm DL software. The median values are automatically written to a spreadsheet.

TABLE 4
Errors in Measuring Terms Used to Calculate Sky Brightness

| | Error (mag) | | |
|---|---|---|---|
| Term | Small-Format | Large-Format | Measurement Method |
| $Q$ ............... | $\pm 0.03$ | $\pm 0.01$ | Standard deviation of the mean of many data sets. |
| $P_{\mathrm{sky}}$ ............. | $\pm 0.02$ max | $\pm 0.03$ max | Deviation of the mean of replicate measures plus an estimate of systematic error. |
| $t$ ................ | Negligible | Negligible | Unknown but estimated to be very small, based on exposure length (12 s typical). |
| $A_{\mathrm{sky}}$ ............. | Negligible | Negligible | Standard deviation of the mean image scale of many data sets, typically less than 0.05%. |

The resulting table includes 5069 records with alt-az coordinates and median ADU pixel$^{-1}$ values. These values are converted to magnitude per square arcsecond using the formula

$$m_{\mathrm{sky}} = Q - 2.5 \log \left( \frac{P_{\mathrm{sky}}/t}{A_{\mathrm{sky}}} \right),$$

where $m_{\mathrm{sky}}$ is sky brightness in magnitudes per square arcsecond, $Q$ is instrumental zero point scaled to 1 s exposure, $P_{\mathrm{sky}}$ is the median sky ADUs pixel$^{-1}$, $t$ is the integration time in seconds, and $A_{\mathrm{sky}}$ is the area of 1 pixel in square arcseconds.

The plate scale, used to determine $A_{\mathrm{sky}}$, is a vital measurement for accurate determination of sky brightness. Plate scales were determined for each camera, based on the astrometric solutions, and represent mean values across the frame. Once again, the wide field of the large-format camera presented a challenge, in that pincushion lens distortion resulted in a variation in scale from center to corner of about 4%. A representative image was divided into 64 subframes, and each subframe was solved separately. The average of the 64 separate plate scale measurements determined $A_{\mathrm{sky}}$ for this camera format.

Each component of the above equation contains both random and systematic measurement errors that can be measured or estimated. These values are shown in Table 4.

Random errors in $Q$ result from poorer extinction regression fits, primarily related to atmospheric effects. Data sets that produced the best fits were used to determine each camera's color transformation term (see Figs. 12 and 13). However, slight systematic variations in $Q$ from night to night may occur for physical reasons, most commonly the accumulation of dust in the optical path. In addition, it is possible that the leaves of the iris of the lens do not seat in exactly the same spot each time the click stop is moved, especially if the f/2 setting is approached from a different direction. Therefore, we attempt to acquire as many data sets per night as possible and take an average of all $y$-intercept values in order to minimize random and systematic errors in the value of this constant as much as possible. With only one data set, no estimate of error can be derived, but the maximum range of variation in $Q$ over many nights (including both random and systematic errors) was observed to be $\pm 0.03$ mag with either format camera.

Systematic error in $A_{\mathrm{sky}}$ may result from using a slightly different focus point from night to night; this can also be fully corrected by applying the calculated average image scale from each night of observations, rather than using a fixed image scale. However, systematic errors in $P_{\mathrm{sky}}$ cannot be removed, as they result from inherent errors in producing a master flat (especially for the large-format system) and the linearizing functions for each camera (described above). Regions near the edges of the field of images (subject to the greatest vignetting by the optical system) will contain more error, as will darker areas of the sky with low ADU pixel$^{-1}$ values, on which the effect of linearizing is more significant. The values given in Table 4 for the error in this variable include an estimate of the maximum systematic error. The total maximum error in sky background brightness measurements for both systems is 4% or less (0.04 mag).

### 2.5.8. Contour Maps of Night-Sky Brightness

Besides providing a convenient and manageably sized data table representing the brightness of the entire sky, the 5069 point grid provides data for the creation of a spline-fit surface model. Visual representation of the data with false color allows at-a-glance comparisons from site to site or over time at the same site. Figures 7 through 9 show examples of this type of graphical display.

### 2.5.9. Indices of Sky Quality Based on Sky Brightness

The following characteristics of the all-sky images may be used as indices of sky quality: (1) surface brightness of the zenith and the darkest and brightest portions of the sky, in mag arcsec$^{-2}$, (2) integrated brightness of the entire sky background (excluding stars), and (3) total integrated brightness of light domes produced by artificial lighting in localized areas, such as cities.

The first index is merely the minimum or maximum value of the clear sky. That is, it is in a region not obscured by foreground objects or clouds and is at least 10° above the horizon, as obscuring haze may sometimes lead to very dark values being measured near the horizon. Values of 21.3 to 22.0 mag arcsec$^{-2}$ in $V_J$ in areas of the sky near a Galactic pole are generally considered to represent natural (unpolluted) conditions (Krisciunas 1997; Walker 1970, 1973). Background sky brightness at the zenith (often the darkest or nearly the darkest part of the sky) is a commonly used index in models that predict the effects of light pollution on sky brightness (Cinzano & Elvidge 2004; Garstang 1986, 1989). In cases where the Milky Way is present near the zenith, these values may be affected by our inability to remove contamination from unresolved stars described above.

In addition to zenith (or darkest) values, the surface brightness of the brightest portion of the sky can be used as a measure of the impairment to natural conditions. That is, a very bright patch of sky near the horizon associated with a city or lighted facility can significantly impair night-sky quality by creating a distraction, illuminating the land from the direction of the light source and casting shadows similar to a moonlit night. We find from field observations that when sky brightness at any point exceeds about 19 mag arcsec$^{-2}$, significant impairment to human night vision and easily noticeable ground illumination occurs. At very dark unpolluted sites, the location of the brightest measurement must be checked to ensure that it reflects natural sky glow rather than a location within the Milky Way.

Total integrated sky brightness can be computed by summing measurements from the 5069 points and multiplying by the number of square arcseconds in 4 deg$^2$ (since each point represents 4 deg$^2$ of sky). This measurement will include a contribution from unresolved stars at low Galactic latitudes, which can be significant at dark sites, particularly when the Galactic center is above the horizon.

Both of the measurement types described above will be affected, particularly in relatively unpolluted sites, by variations in natural sky brightness arising from changes in natural airglow and atmospheric aerosols. Variations in these contributions can occur on timescales from minutes to years. At sites with significant blocking of the sky by trees or terrain features, the total integrated sky brightness will be affected. Therefore, direct comparisons of these numbers from one data set to another must be made with caution.

The third index, total magnitude of an individual city's light dome, can be computed in a similar manner to the total integrated sky brightness by restricting the summation to only those values that correspond to sample points within the light dome (see Fig. 5). This method is still under development, and the results presented below in § 3.3 are considered preliminary. "Natural" sky brightness must be subtracted from each point's value in order for the net contribution of the city's light pollution to be accurately quantified. In addition, a 2° grid may be too coarse to produce an accurate measurement where brightness gradients are steep, in particular close to the horizon. A finer sampling grid within the light dome, or a total summation of pixels within the light dome after applying a median filter to remove stars, may provide a better technique. Light domes fade gradually toward the natural background as the angle from the light pollution source increases; to simplify the approach, we defined light domes as those areas of the sky that exhibited more than 2 times the expected natural sky back-ground for a given altitude above the horizon, as determined from a model of natural sky brightness with altitude above the horizon. In Figure 5, sample points on the 2° grid are shown; those points falling within the bold white outline are 2 times or more the brightness of the modeled sky background. Light domes that overlap the Milky Way or zodiacal light were avoided for the purpose of this analysis.

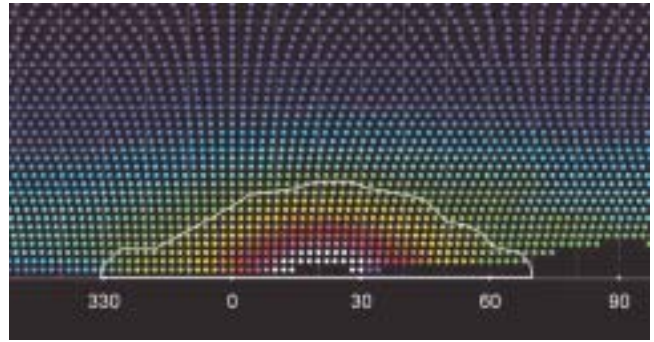

Fig. 5.—Light dome of Las Vegas, Nevada, as seen from Mojave National Preserve, and its boundary (*white outline*) defined as described in the text. The numbers shown at the bottom represent the azimuth in degrees. The color of each sample point represents sky brightness, following the legend in Figs. 7–9.

## 3. RESULTS

### 3.1. Selected All-Sky Data

To date, we have collected data from over 80 different locations in more than 45 areas administered by the National Park Service, as well as from the US Naval Observatory (USNO) Flagstaff Station, Lowell Observatory, and Palomar Observatory. Here we present data from the following locations: Death Valley National Park, California; Mojave National Preserve, California; Lake Mead National Recreation Area (NRA), Nevada-Arizona; Sunset Crater National Monument (NM), Arizona; Walnut Canyon NM, Arizona; Wupatki NM, Arizona; and USNO Flagstaff Station, Arizona. These sites are grouped around two cities, Las Vegas, Nevada, and Flagstaff, Arizona, and are shown on a map of the southwestern US in Figure 6. In the discussion that follows, all magnitude values are on the $V_{\mathrm{NPS}}$ system, and except as noted are in units of magnitudes per square arcsecond.

Sky brightness data from these sites are presented in Table 5, and fish-eye representations of the data as sky brightness maps in false color are shown in Figures 7 through 9. Figure 7 displays data from three different sites in Lake Mead NRA, near the city of Las Vegas, Nevada. Overton Beach (Fig. 7*a*) and Temple Bar (Fig. 7*b*) are relatively dark sites, located 79 and 83 km from the center of Las Vegas, and with the darkest portion of the sky measured at 21.51 and 21.76 mag arcsec$^{-2}$ (Table 5), respectively. Nevertheless, the light dome of Las Vegas/Henderson/Boulder City dominates the western (right) portion of the sky at each location. Smaller light domes from two additional cities can also be seen in Figure 7a, and the winter Milky Way causes the light band that appears from left to right across the sky. The summer Milky Way is apparent in Figure 7b across the eastern (left) part of the sky, with brighter portions to the south (bottom). Figure 7c represents the sky brightness at Government Wash, only 36 km from Las Vegas, although within the boundaries of the National Recreation Area, virtually a suburban site. Dramatically brighter, light pollution dominates the sky, overwhelming natural features. The darkest value measured was 20.12 mag arcsec$^{-2}$.

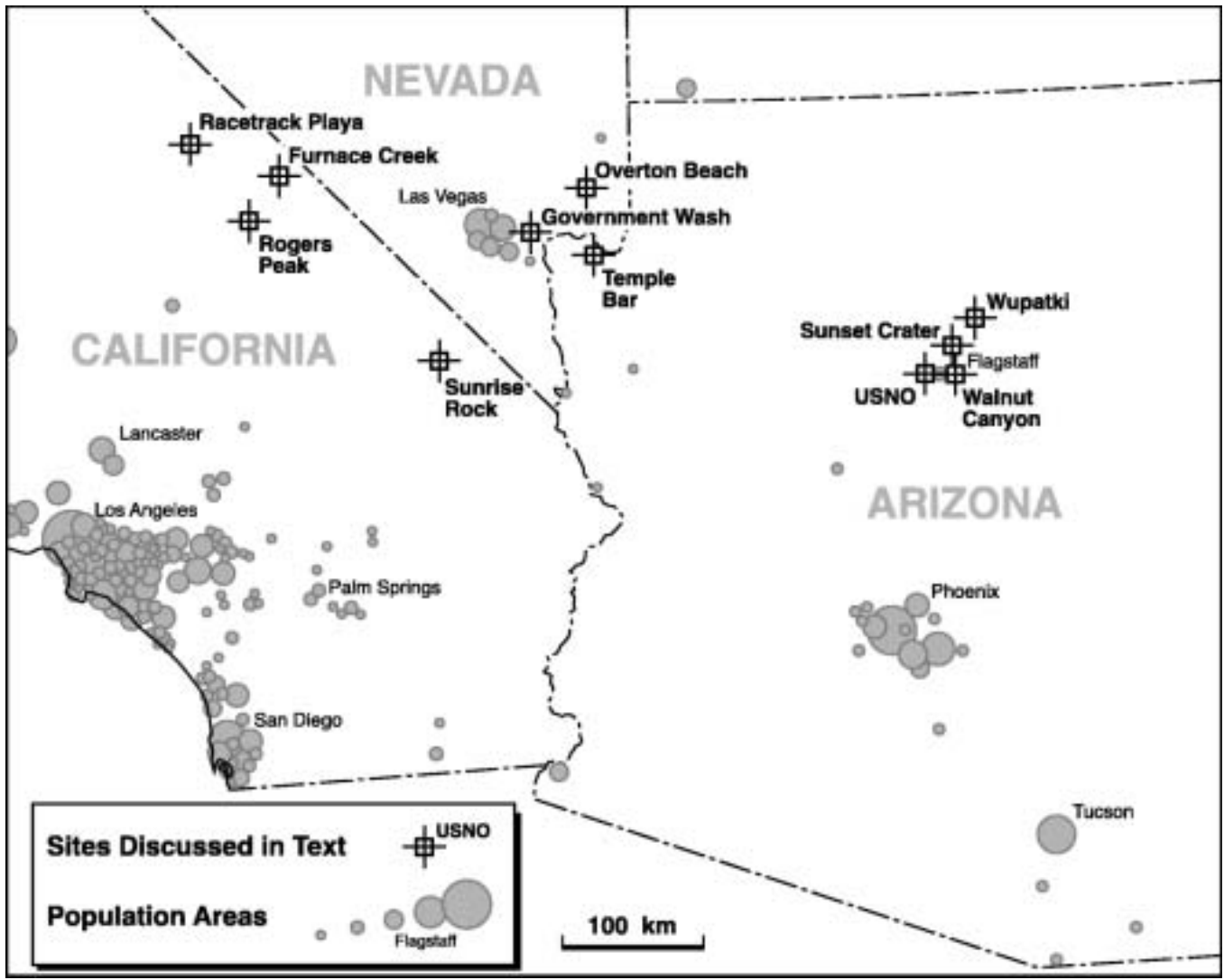

FIG. 6.—Map of the southwestern US showing the location of monitoring sites relative to major cities.

Relatively unpolluted night skies are represented in Figure 8, from Death Valley National Park and Mojave National Preserve, California. Figures 8a and 8b are from the same location, Furnace Creek in Death Valley, on two different nights, and illustrate the wide variation in natural airglow. In

Figure 8a, the sky background is very dark at 22.12 mag arcsec$^{-2}$, and features of the zodiacal light and Milky Way can be seen in the all-sky map. The broad, triangular-shaped bright area of light and dark blue to the west (right) extending nearly to the zenith is the zodiacal light, the nearly straight band of light extending from lower left to upper right is the winter Milky Way, while bright bands of light skirting the horizon to the south (bottom) and east (left) are light domes from southern California cities and Las Vegas, respectively. The gegenschein is barely visible to the left of center, about halfway from horizon to zenith. In Figure 8b, bright airglow produces a background sky brightness of 21.68 mag arcsec$^{-2}$, 0.44 mag brighter than in Figure 8a (see Table 5), and washes out features of the zodiacal light and winter Milky Way. The light dome of Las Vegas is represented in data from Dante's View (135 km distant; Fig. 8c), Rogers Peak (168 km distant; Fig. 8d), and

TABLE 5
Sky Brightness Indices from Selected Sites, in $V_{NPS}$

| Location | Altitude (m) | Date, UT Time | Extinction Coefficient (mag air mass$^{-1}$) | Darkest Sample Point | | | Brightest Sample Point | | | Integrated Sky | |
|---|---|---|---|---|---|---|---|---|---|---|---|
| | | | | (mag arcsec$^{-2}$) | Az (deg) | Alt (deg) | (mag arcsec$^{-2}$) | Az (deg) | Alt (deg) | Total (mag) | Alt ≥ 20° (mag) |
| Lake Mead NRA: | | | | | | | | | | | |
|   Overton Beach ............ | 459 | 2004 Dec 14, 03:59 | 0.153 | 21.51 | 257 | 72 | 17.37 | 244 | 4 | −7.86 | −6.99 |
|   Temple Bar ................ | 517 | 2005 May 11, 08:23 | 0.178 | 21.76 | 36 | 74 | 16.80 | 282 | 2 | −7.86 | −6.91 |
|   Government Wash ........ | 423 | 2004 Dec 3, 03:14 | 0.166 | 20.12 | 74 | 76 | 15.62 | 272 | 2 | −9.78 | −8.65 |
| Death Valley NP: | | | | | | | | | | | |
|   Furnace Creek ............ | 137 | 2004 Feb 11, 04:42 | 0.167 | 22.12 | 35 | 46 | 20.00 | 196 | 2 | −6.86 | −6.33 |
| | | 2004 Dec 12, 07:04 | 0.152 | 21.68 | 17 | 54 | 20.07 | 194 | 2 | −7.23 | −6.68 |
|   Dante's View .............. | 1675 | 2004 May 23, 06:24 | 0.170 | 21.87 | 60 | 88 | 18.40 | 94 | 2 | −7.32 | −6.89 |
|   Rogers Peak .............. | 3047 | 2004 May 23, 07:12 | 0.132 | 21.95 | 354 | 70 | 18.66 | 92 | 2 | −7.26 | −6.62 |
| Mojave National Preserve: | | | | | | | | | | | |
|   Sunrise Rock .............. | 1535 | 2004 Nov 18, 06:48 | 0.133 | 21.53 | 280 | 74 | 17.46 | 22 | 2 | −7.66 | −6.87 |
| Sunset Crater NM: | | | | | | | | | | | |
|   Lava Flow Trailhead ...... | 2133 | 2005 Feb 9, 06:43 | 0.149 | 22.16 | 18 | 48 | 18.01 | 196 | 2 | −6.96 | −6.38 |
| Wupatki NM: | | | | | | | | | | | |
|   Wukoki Ruin .............. | 1411 | 2004 Jun 12, 07:32 | 0.196 | 22.25 | 295 | 76 | 20.17 | 216 | 4 | −7.09 | −6.54 |
| Walnut Canyon NM: | | | | | | | | | | | |
|   West Rim View ........... | 2045 | 2004 Jun 19, 06:37 | 0.161 | 21.87 | 25 | 76 | 18.50 | 282 | 2 | −7.38 | −6.72 |
| USNO Flagstaff Station: | | | | | | | | | | | |
|   West Parking Lot ......... | 2315 | 2004 Sep 14, 07:57 | 0.131 | 21.91 | 253 | 78 | 19.44 | 82 | 6 | −7.07 | −6.59 |
| | | 2005 Feb 9, 07:43 | 0.123 | 21.71 | 335 | 76 | 18.55 | 196 | 2 | −7.43 | −6.85 |

Sunrise Rock (103 km distant; Fig. 8e). These data are discussed below.

Locations near the city of Flagstaff, Arizona, are shown in Figure 9. Sunset Crater NM (21 km distant; Fig. 9a) and Wupatki NM (45 km distant; Fig. 9b) both display dark sky background at 22.16 and 22.25 mag arcsec$^{-2}$, respectively. Examination of the all-sky maps reveals that the Flagstaff light dome is more prominent at Sunset Crater, while at Wupatki, a brighter natural airglow "ring" is apparent near the horizon on the night the observations were made. Testament to the success of light pollution controls within the city of Flagstaff is evidenced in the data from Walnut Canyon NM (Fig. 9c), only 11 km from the center of Flagstaff. While a broad dome of light from Flagstaff is evident to the west, the zenith sky brightness is very dark, measured at 21.87, and considerable detail is evident in the summer Milky Way, rising in the east (left). Two data sets were selected for comparison from the US Naval Observatory Flagstaff Station, one in late summer (Fig. 9d) and one in winter (Fig. 9e). Both were taken near 01:00 A.M. local time and with the same equipment. Sky brightness numbers for the darkest portion of the sky were quite dark (21.71 for winter, 21.91 for summer), despite being located only 8 km from the center of Flagstaff (population approximately 55,000). It should be noted here that an important characteristic of the Flagstaff lighting code is a strong requirement for the use of low-pressure sodium, which has a reduced effect on the visible and $V_{NPS}$ sky glow, as discussed in § 4.1. All other indices from the winter set were slightly brighter than the summer, which would be expected because of reduced obstruction of lighting due to lack of foliage on broadleaf trees within the city, as well as the presence of a small amount of reflecting snow on the ground at Flagstaff.

Table 5 lists sky brightness indices described in this paper for each of these sites. These include darkest sample point, brightest sample point, the integrated brightness of the whole sky, and the integrated brightness of the sky above 20° elevation. The value for darkest sample point (this typically would be near the zenith, unless the Milky Way is overhead) can be compared with the theoretical natural "baseline" for unpolluted skies, about 22.0 mag arcsec$^{-2}$ in $V_{NPS}$. Most of the sites investigated are near that value, except Government Wash at Lake Mead.

Values for the brightest portion of the sky are of interest to the National Park Service because they represent unnatural intrusions on the nightscape, will prevent full human dark adaptation, and may have effects on wildlife. Our observations of the brightness of the brightest star clouds in the Milky Way indicate that values in the range 19.5–20.0 represent a typical maximum brightness for this feature under dark skies. Any areas of the sky that are measured as brighter than this may be considered "unnatural." It can be seen from Table 5 that several National Park Service sites that are generally considered fairly remote produced values brighter than 19.5. Namely, Dante's View (18.40), Rogers Peak (18.66), Lava Flow Trailhead in Sunset Crater (18.01), Sunrise Rock (17.46), and all locations surveyed in Lake Mead National Recreation Area, including the suburban Government Wash, at 15.62.

The integrated brightness of the entire sky background (excluding stars and planets) is an excellent index of sky quality, as it is a quantity that is site-specific and has significant relevance to the human visual experience. Of the sites shown in

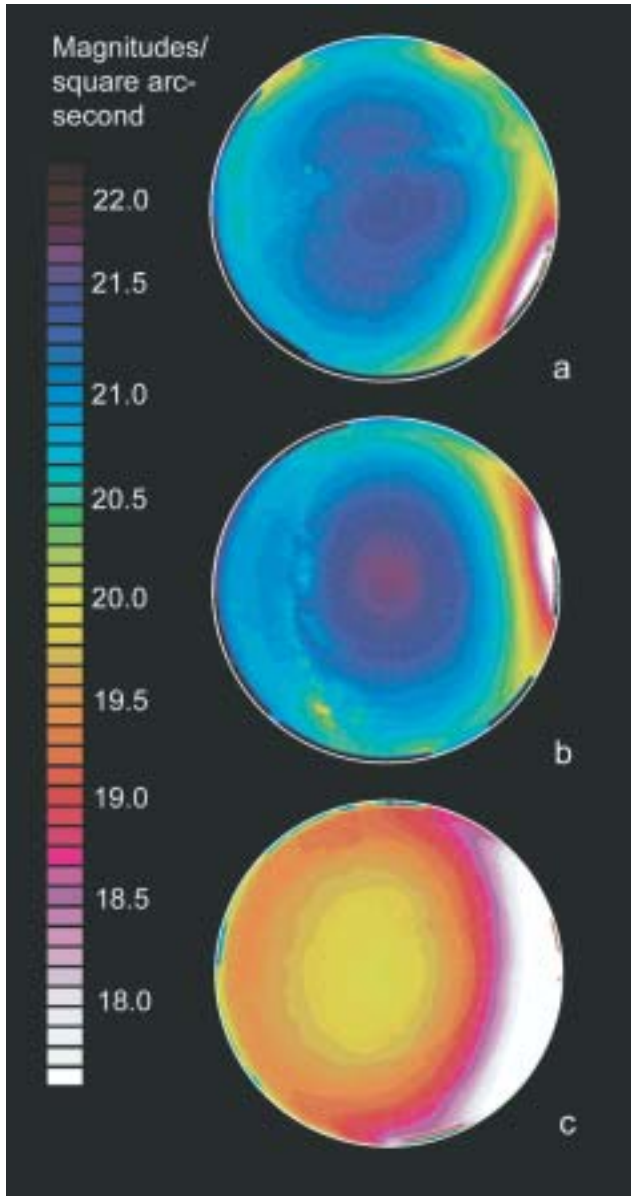

FIG. 7.—Fish-eye representations of sky brightness data from Lake Mead NRA: (*a*) Overton Beach, (*b*) Temple Bar, and (*c*) Government Wash. North is up, east to the left, and the zenith is at the center of each map.

Table 5, Lava Flow Trailhead in Sunset Crater and Furnace Creek (2004 February 11) in Death Valley were measured as the darkest, at magnitude $-6.96$ and $-6.86$, respectively. This is probably due in part to significant foreground obstructions (the profile of Sunset Crater itself and Lennox Crater obscured the view at Lava Flow Trailhead, while Furnace Creek is in a deep valley ringed by mountains). Therefore, total sky brightness observations from Sunset Crater would be expected to be darker than a mountaintop site with unobstructed views to the horizon. Such obstructions were a strong factor at the USNO

Flagstaff Station (summer), where nearby trees and the dome of the 1.55 m telescope obscured large portions of the sky. This site produced an uncharacteristically dark value, of magnitude $-7.07$, for total sky brightness, despite the obvious light dome from Flagstaff on the eastern horizon. The effects of the light dome of Las Vegas are evidenced in this statistic for the Lake Mead sites, with Government Wash registering a very bright magnitude $-9.78$, about as bright as the Moon when 50% illuminated.

By excluding the portion of the sky within 20° of the horizon, effects of foreground obstructions are reduced and a more representative measure of sky brightness emerges. Government Wash remains the brightest sky in our sample, at magnitude $-8.65$, but Furnace Creek, at magnitude $-6.33$, barely surpasses Lava Flow Trailhead ($-6.38$) as the darkest. As more data sets are collected, binning these values into qualitative categories may provide a useful descriptive tool.

### 3.2. Atmospheric Extinction

Table 5 lists the extinction values measured for the 11 data sets. They range from 0.123 to 0.196 mag air mass$^{-1}$. These variations are attributable to variations in total amount of atmosphere above the observing site (i.e., altitude) and total aerosol content of the atmosphere, varying both due to local conditions such as weather, dust, and air pollution, and generally by season (see Krisciunas 1987). Light pollution sources may be amplified or reduced by high-extinction conditions, depending on the distance from the source to the observer (Garstang 1986). Most of the sites discussed in this paper are at some distance from the sources, and higher extinction values generally produce darker skies because of the attenuation of light through many kilometers of sky brightness. Natural sources of sky brightness, especially the airglow, may be accentuated by very transparent skies at higher altitudes. In future, as more data sets with different extinction values are obtained from a given site, a quantitative analysis of the effect of this factor on observed sky brightness may be attempted.

### 3.3. Total Integrated Light from Las Vegas, Nevada, Light Dome

The brightness of the light dome of Las Vegas, Nevada, was measured using data from several monitoring sites in Lake Mead National Recreation Area, Nevada-Arizona, Death Valley National Park, California-Nevada, and Mojave National Preserve, California. In each case, a model of "natural" sky brightness with altitude was constructed by selecting a portion of the data set that excludes obvious light domes from cities and towns, bright parts of the Milky Way, and the zodiacal light. At Government Wash, the entire sky is brightened by the lights of Las Vegas, and no azimuth offers an unpolluted profile. For this data set, a natural profile extracted from the darker sites was used. The modeled natural sky brightness was subtracted from the observed sky brightness values within the light dome.

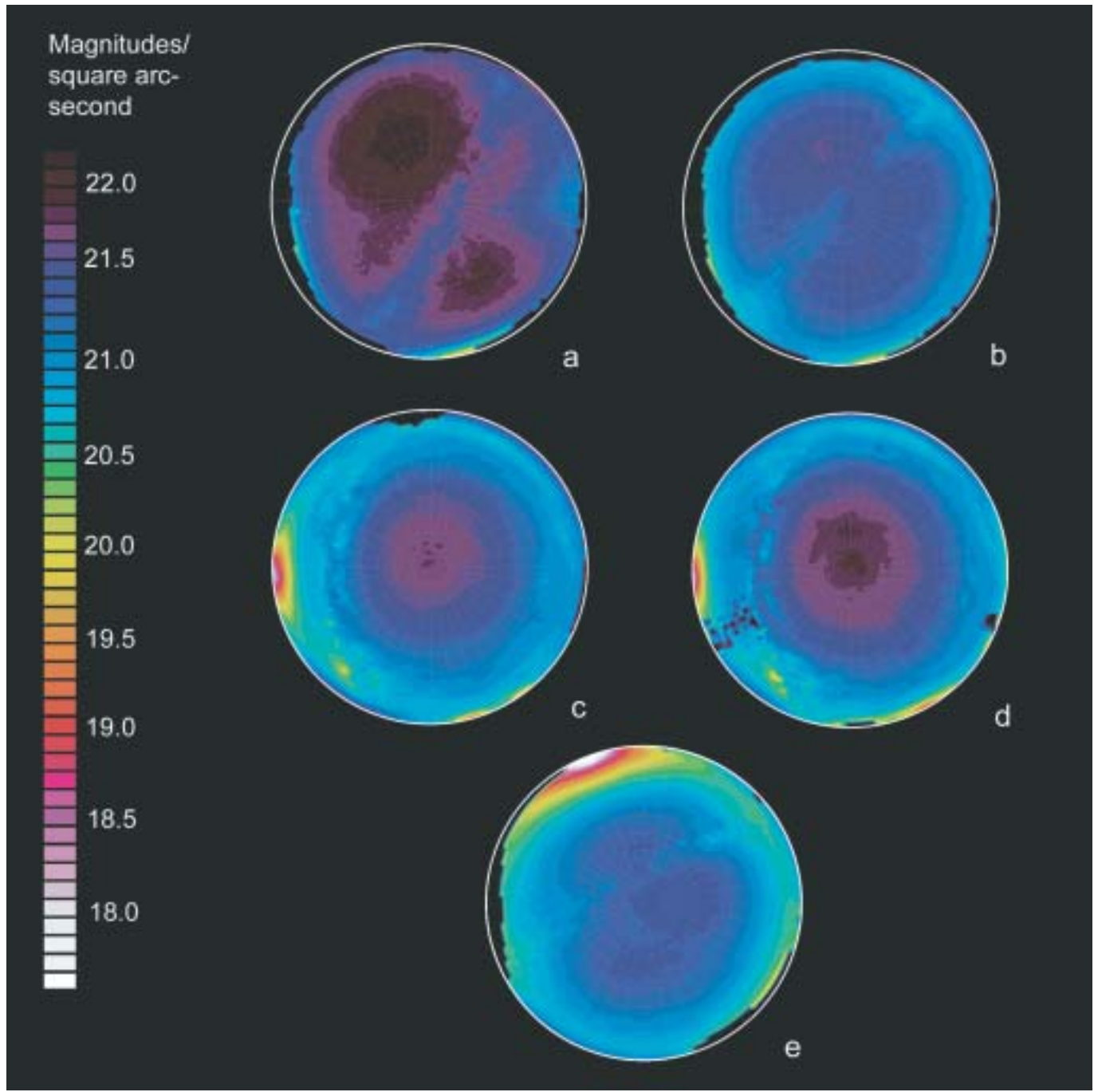

FIG. 8.—Fish-eye representations of night-sky brightness data from Death Valley National Park and Mojave National Preserve: (*a*) Furnace Creek, 2004 February, (*b*) Furnace Creek, 2004 December, (*c*). Dante's View, (*d*) Rogers Peak, and (*e*) Sunrise Rock. North is up, east to the left, and the zenith is at the center of each map.

Table 6 shows the distance from Las Vegas, the integrated light dome brightness, and the total light dome area in square degrees of sky at four sites where the view of the Las Vegas light dome is relatively unobscured.

At every site except Rogers Peak in Death Valley National Park, the total light produced by the city sky glow is brighter than any permanent celestial object except the Moon. This indicates that one must travel between 100 and 170 km from the city center to arrive at a sky that approximates "natural" light. Even at Rogers Peak, Las Vegas produces more light than

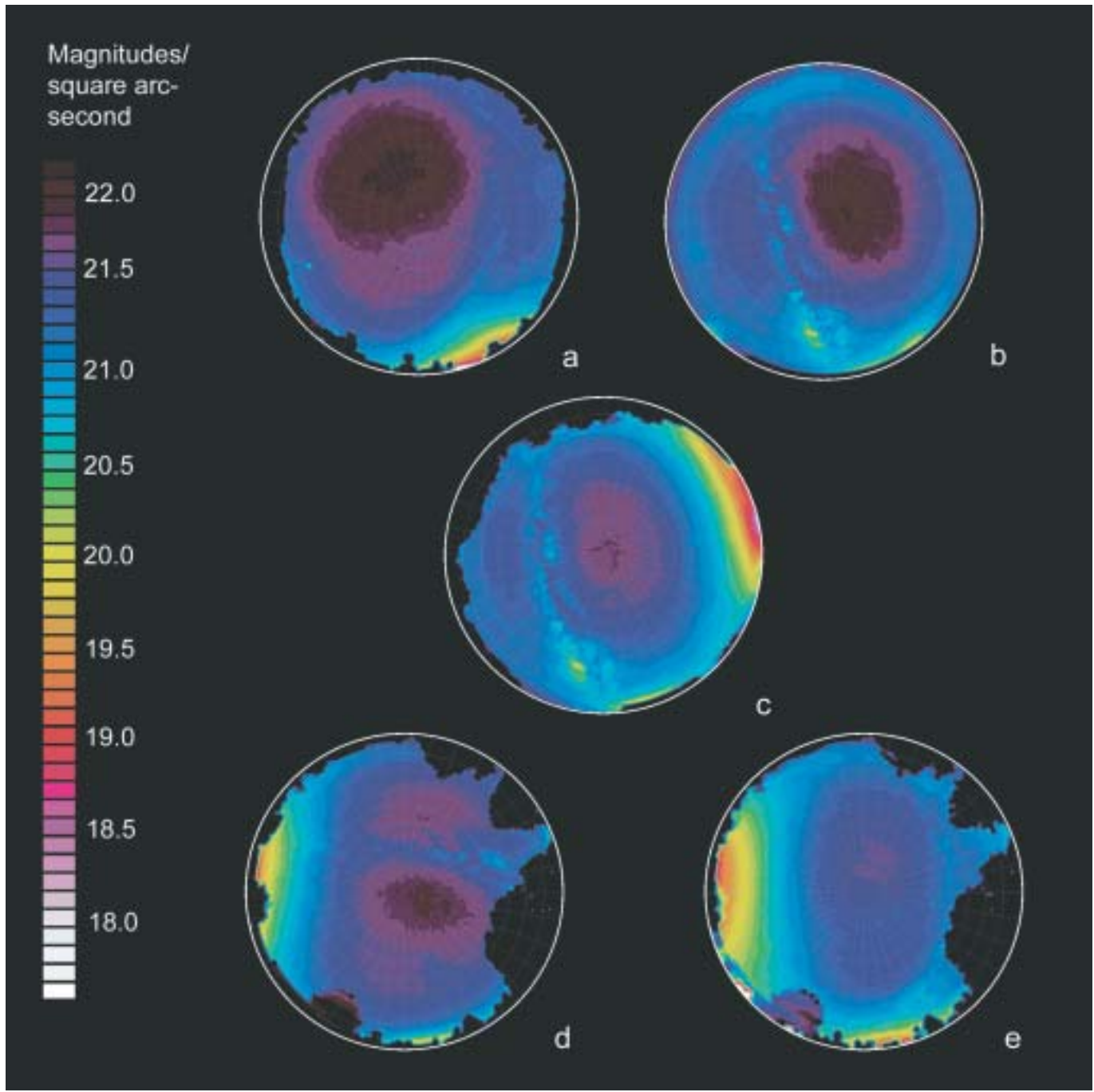

FIG. 9.—Fish-eye representations of night-sky brightness data from sites near Flagstaff, Arizona: (*a*) Lava Flow Trailhead, Sunset Crater NM, (*b*) Wukoki Ruin, Wupatki NM, (*c*) West Rim View, Walnut Canyon NM, (*d*) USNO Flagstaff Station, 2004 September, and (*e*) USNO Flagstaff Station, 2005 February. North is up, east to the left, and the zenith is at the center of each map.

the planet Jupiter. At the Government Wash site in Lake Mead National Recreation Area, the city's light comprises more than half of the light arising from the entire sky. Least-squares regression to the closest three points (Government Wash, Temple

Bar, and Sunrise Rock) yields a slope of $-2.50 \pm 0.07$ (see Fig. 10), precisely reproducing the brightness-distance relation described by Walker (1977), although these data refer to a total flux rather than sky brightness differences as modeled by Wal-

| Location | Date, UT Time | Distance from Las Vegas (km) | Light Dome Brightness (mag) | Light Dome Area (deg$^2$) |
|---|---|---|---|---|
| Rogers Peak, Death Valley NP ………… | 2004 May 23, 07:12 | 173.2 | −3.17 | 132 |
| Sunrise Rock, Mojave NP ……………… | 2004 Nov 18, 05:50 | 101.7 | −5.61 | 900 |
| Temple Bar, Lake Mead NRA ………… | 2005 May 11, 08:23 | 76.7 | −6.47 | 1676 |
| Government Wash, Lake Mead NRA …… | 2004 Dec 3, 03:14 | 29.1 | −9.03 | 4224 |

ker. That the most distant point falls below this line is likely due to the Earth's curvature complicating the light path (Walker 1977).

## 4. DISCUSSION

### 4.1. System Photometric Characteristics

The NPS photometric system has been designed to measure sky brightness in a band that closely approximates Johnson $V$, the response most widely used in published astronomical literature on sky brightness. This allows our measurements to be compared with the work of others, and furthermore, is a reasonable approximation to human visual response and is therefore valuable for evaluating the impact of light pollution for visitors to national parks. But this system, like any broadband photometric system, has disadvantages as well. The source of these problems lies in the fact that instrumental zero point and color terms used to relate any given individual system to the standard system are based on measurements of standard stars, which exhibit spectra approximately characterized by blackbodies of various temperatures. As the visible portion of the night-sky spectrum is not similar to a stellar spectrum and, furthermore, varies significantly in both time and position on the sky, measurements of night-sky surface brightness made with modern CCD systems, or indeed any system not identical to the standard system, cannot reliably be transformed to the $V_J$ system, because the responses of such systems are inevitably to some degree different than the $V_J$ response. Because of these effects, some of which are discussed in more detail below, all-sky brightness measurements presented in this paper are on the native NPS system, i.e., they have had no color term corrections applied. In the following, we explore both the consequences of differing system spectral responses, as well as the effects arising from the peculiar spectral characteristics of the natural and light-polluted night sky.

To help simplify and clarify the ensuing discussion, the terminology listed in Table 7 will be used to describe the several spectral responses and magnitude systems.

#### 4.1.1. CCD Comparison

First, we examine the responses of the two detector/filter systems used, the KAF 261E and KAF 1001E with a Bessell $V$ filter, used in the small- and large-format cameras, respectively. Figure 11 shows the relative response of the two camera systems, normalized for equal area under the curves. Table 8 shows the effective wavelengths and full widths at 50% and 10% of peak response. Overall, the responses are very similar, particularly in the important shoulder regions, although there are up to 20% deviations in sensitivity in the midrange. This may lead to significant deviations in measurements using the two systems when particular emission lines, such as [O I] 557.7 nm, dominate the spectrum, as may occur under dark sky conditions with strong airglow, although other strong emission features arising from artificial light sources (such as sodium and mercury vapor [MV]) will be detected similarly with the two camera systems. Under light-polluted conditions, we expect sky brightness measurements made with the two systems to agree within a few percent.

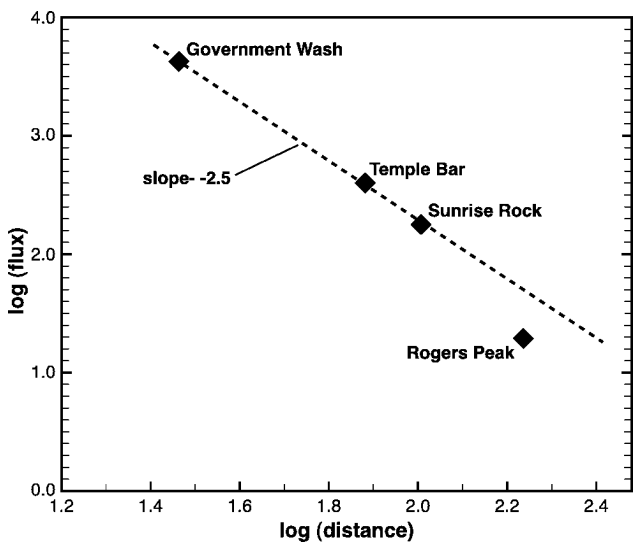

FIG. 10.—Integrated brightness of the Las Vegas light dome vs. distance from Las Vegas. The dashed line is described in the text.

TABLE 7
SPECTRAL RESPONSE AND MAGNITUDE
TERMINOLOGY

| Spectral Response | Magnitude |
|---|---|
| Johnson $V$ …………… | $V_J$ |
| Johnson $B$ …………… | $B_J$ |
| NPS ………………… | $V_{NPS}$ |
| (Human) photopic …… | $M_p$ |
| (Human) scotopic …… | $M_s$ |

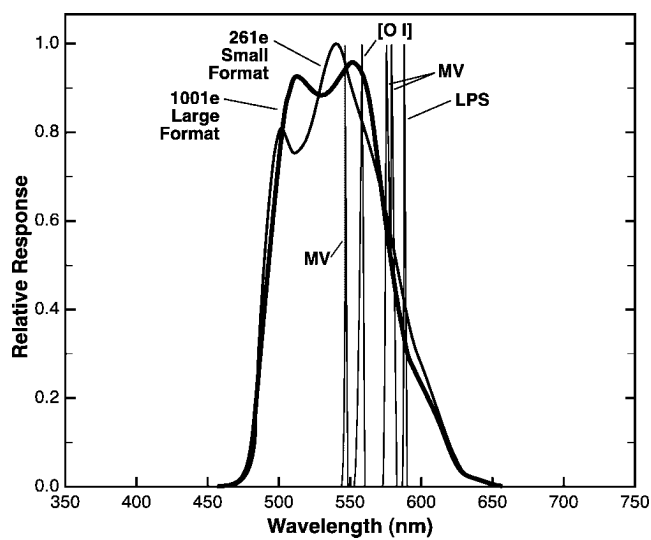

Fig. 11.—Relative response of the KAF 261E and KAF 1001E detectors combined with the Bessell *V* filter. Important emission lines of MV (Hg), [O I], and LPS (Na) are shown.

### 4.1.2. Instrumental Zero Point and Color Terms

Instrumental zero points and color terms were set for each camera by determining the relation between $V_J - v$, air mass, and standard $(B_J - V_J)$ color for standard-star measurements ($V_J$ and $[B_J - V_J]$ as published in the *Hipparcos* catalog). A number of different nights with stable photometric conditions were analyzed for each camera. Figure 12 shows a representative color relation for the SBIG camera. A linear regression fits these data quite well, indicating that higher order terms are not required. After the color transformation terms were determined for each camera, they were applied to each standard-star measurement, then the regression with air mass was recalculated to determine extinction coefficient and instrumental zero point. Examination of $(V_J - v)$ versus air mass after color effects have been removed reveals not only the extinction coefficient for each particular data set, but also the instrumental zero point and the precision of the standard-star photometric measurements. Figure 13 shows a typical air mass versus $(V_J - v)$ plot. A value of 0.025–0.035 for a standard error of $v$ and 0.01 for a standard error of the instrumental zero point are typical for the large-format system.

Table 2 lists the zero points and color terms determined for each camera. We expect and observe that these values are generally stable. Nonetheless, these terms are monitored by per-forming an unconstrained regression on each data set, and significant changes trigger an inspection of the camera and optics. The extinction coefficients derived from multiple data sets on a particular night are calculated using a fixed instrumental zero point for all sets, assuming no changes will occur to the instrument over the course of a single night. Occasionally, significant changes have been observed but were attributed to either high clouds moving through or frost forming on the front surface of the optics. Data sets influenced by such environmental factors are discarded.

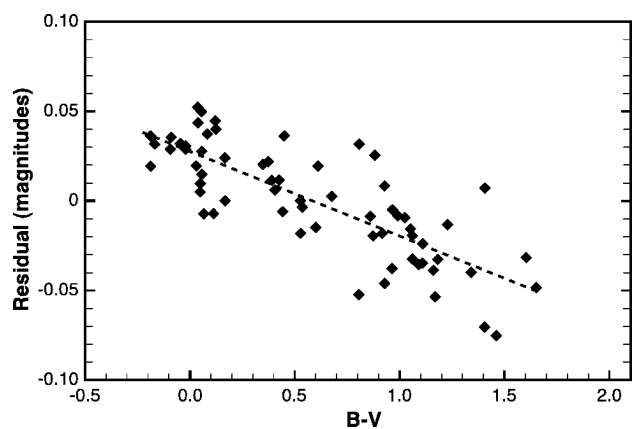

Fig. 12.—Plot of standard stars $(B_J - V_J)$ vs. residuals of a linear fit of air mass vs. $V_J - v$ for the SBIG camera. The line of slope −0.048 mag per magnitude of $(B_J - V_J)$ represents the color transformation term between stellar photometry with this camera and photometry in the Johnson *V* band as published in the *Hipparcos* catalog.

### 4.1.3. Sky Glow/Artificial Lighting Spectral Effects

Since the spectral responses of the $V_{NPS}$ and $V_J$ systems differ slightly, and since color terms defined by observations of stars cannot be used to transform measurements of sky brightness onto the $V_J$ system, small color effects persist in the NPS sky brightness observations. Since the effective wavelength of the $V_{NPS}$ response is about 10 nm bluer than $V_J$, two effects are expected. The first is that the $V_{NPS}$ natural (unpolluted) sky brightness will be somewhat darker than the value $V_J \sim 22.0$ commonly attributed to an unpolluted sky by Garstang (1986) and others, since the natural sky spectrum (Fig. 14) decreases in brightness toward the blue (consistent with the observation that measurements made of a natural sky in $B_J$ approach magnitude 23.0). A simple linear interpolation would lead to a prediction of about 0.1 mag fainter, which is verified by the analysis and observations described below.

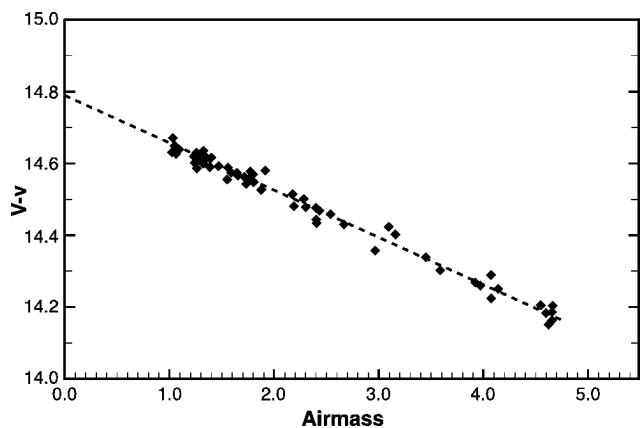

Fig. 13.—Example "extinction plot" showing offset of measured magnitudes from standard system as a function of air mass after the color transformation term is applied.

TABLE 8
Effective Wavelengths and Full Widths through the
Bessell *V* Filter

| Detector | $\lambda_{eff}$ | Full Width (nm) 50% | 10% |
| --- | --- | --- | --- |
| KAF 261E ............... | 541 | 93 | 136 |
| KAF 1001E .............. | 541 | 87 | 136 |

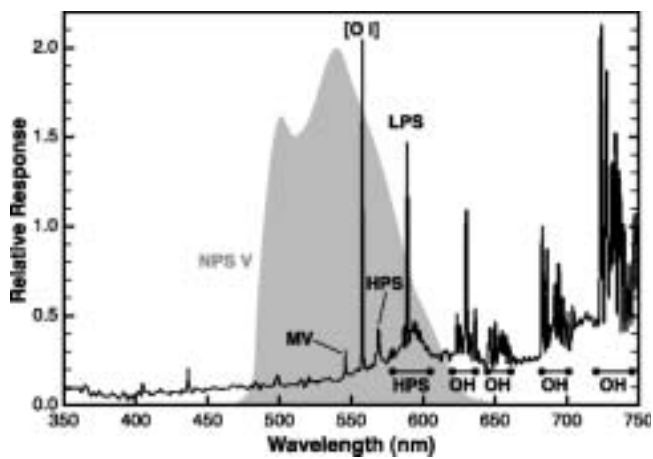

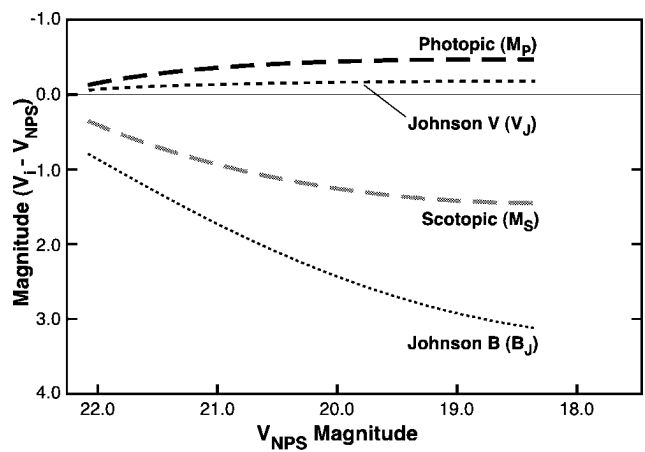

FIG. 14.—Night-sky spectrum from Mount Hopkins, Arizona, 1992 August, with response of the NPS $V$ band superimposed.

FIG. 15.—$V_{NPS}$ vs. $V_J$ (*short dashed line*), $B_J$ (*dotted line*), $M_p$ (*black dashed line*) and $M_s$ (*gray dashed line*) magnitude offsets for a sky polluted with high-pressure sodium emission.

The second effect is that the offset between $V_{NPS}$ and $V_J$ will vary as light pollution levels increase and as the mixture of light sources contributing to the sky glow changes. The decreased red sensitivity of $V_{NPS}$ (e.g. Fig. 4) in particular makes measurements on this system less sensitive to the emissions of high- and low-pressure sodium lamps. We consider this a positive attribute of $V_{NPS}$, since it makes it somewhat more "scotopic" than $V_J$, i.e., more like the visual response of the eye adapted under dark-sky conditions. In the discussion that follows, this effect is examined in detail for the ideal cases of sky glow caused by pure high-pressure sodium, metal halide, and low-pressure sodium lighting.

Figure 14 shows an example spectrum of the night sky obtained from observations made at the MMT on Mount Hopkins in southern Arizona as part of previous research by one of the authors (C. B. L.) in 1992 August. In this spectrum, taken near sunspot maximum, are evident both the characteristics of natural sky glow, such as the general continuum increasing toward the red, hydroxyl emission in the red, and the strong airglow line from the forbidden oxygen transition at 557.7 nm, as well as a significant contribution from high- and low-pressure sodium lighting primarily from the Tucson area. Detailed examination of the spectrum also reveals evidence of a small amount of mercury vapor emission. By examining uncontaminated spectral features for high-pressure, low-pressure, and mercury vapor lamps, spectra of these lamp types were subtracted from the sky spectrum to approximate a natural "dark sky" spectrum. High-pressure sodium alone was then incrementally added back to this spectrum to simulate a sky polluted by increasing amounts of pure high-pressure sodium lighting. The resulting spectra were convolved with the five responses $V_{NPS}$, $V_J$, $B_J$, scotopic, and photopic. To establish a zero point, it was assumed that the artificial "dark sky" spectrum generated at the beginning of this process yielded a $V_J$ magnitude of 22.0. The same zero point was then applied to measurements on all five systems after their spectral response curves were normalized for equal area. Although the "natural" brightness of the sky at this particular time could likely have been brighter than 22.0, the exact zero point chosen in the following analysis is unimportant, as we seek to examine differences in the fluxes measured by the different responses.

Figure 15 shows these offsets as a function of $V_{NPS}$. The $V_J$ band for all sky brightnesses from natural to strongly light polluted will measure brighter than $V_{NPS}$, ranging from about 0.07 mag under the darkest natural conditions to about 0.18 mag in the brightest skies. This effect is somewhat larger for the photopic response, where the sky will appear about 0.16 mag brighter than $V_{NPS}$ in the darkest conditions (where the photopic response is not actually relevant), rising asymptotically to 0.48 mag under bright conditions. Nonetheless, both the $V_J$ and photopic systems are reasonably closely represented by measurements on the $V_{NPS}$ system. It should be noted that from a visual perspective, a difference of 0.48 mag would be only marginally perceptible to most observers.

Perception of sky brightness by human scotopic vision is important to the National Park Service objective of preserving pristine nighttime scenery. Figure 15 illustrates an important effect where the brightness of a high-pressure sodium-polluted sky perceived by the scotopic eye (appropriate for relatively dark conditions fainter than about 20th mag arcsec$^{-2}$) will appear from 0.4 to 1.4 mag fainter than reflected by the $V_{NPS}$ measurements, a significant and perceptible difference. This implies that the use of lamps of this type for facility lighting near these wild and remote locations will significantly mitigate impairment to the natural experience of a visitor seeking a pristine night sky. The locus of scotopic measurements lies about halfway between the $V_{NPS}$ system and $B_J$. This effect is further strongly dependent on the spectral characteristics of the light pollution source. Figure 16 shows the predicted sky brightness increases expected for high-pressure sodium (HPS), metal halide, and low-pressure sodium (LPS) when equal amounts of "visible" light (defined by the number of lumens added to the natural night-sky spectrum, where lumens are in turn defined by the photopic response) are added to the natural night-sky spectrum. These relations were derived by normalizing the lamp spectra such that the areas under each spectrum were equal when convolved with the photopic response, then incrementally adding these normalized spectra to the natural dark sky spectrum described above and determining the sky brightness magnitudes appropriate for the $V_{NPS}$ and scotopic responses.

It is clear from Figure 16 that when considering the response of the dark-adapted eye, on a lumen-for-lumen basis metal halide lighting brightens the sky considerably more than either HPS or LPS lighting, with metal halide sky glow appearing up to 1.5 mag (4 ×) brighter than HPS under light-polluted conditions, and 1.9 mag (5.8 ×) brighter than LPS. Another way of characterizing this difference is that it takes 4000 lm (lumens) of high-pressure sodium lighting to cause the same amount of visible sky glow as 1000 lm of metal halide lighting. This effect results entirely from the spectral shift of the dark-adapted response of the eye and its relative insensitivity to yellow and red light.

However, this treatment is incomplete, since this visual effect will be compounded by atmospheric effects whereby the bluer metal halide lighting will be more strongly scattered than HPS or LPS light, especially in clear atmospheric conditions appropriate for most observatory sites and national park units in the western US. This effect will lead to a tendency for bluer light to cause greater light pollution when viewed from within the city or nearby, and less when viewed from a great distance. The crossover point will depend on the atmospheric extinction, details of the aerosol content, and other factors, and will not be explored in detail here.

### 4.1.4. Natural Sources of Light

#### 4.1.4.1. Unresolved Stars

The size of the pixels on the sky in both camera systems is large enough that the contribution from unresolved stars will often be nonnegligible, particularly at low Galactic latitudes.

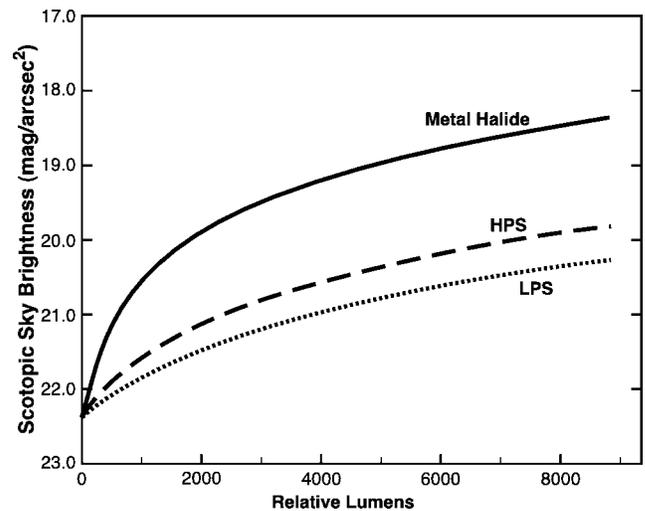

Fig. 16.—Sky brightness vs. relative lumen amounts of artificial lighting in the scotopic system for HPS (*solid line*), metal halide (*dashed line*) and LPS (*dotted line*). Zero point is as described in the text.

That this contamination survives our median filtering process is obvious from the appearance of the Milky Way in Figures 7–9 (this is another advantage to our panoramic measurement system: such contaminated measurements are much harder to recognize in data sets composed of isolated measures). Since the small-format cameras had to be defocused to produce acceptable photometric precision on standard stars, the light from each stellar image is reduced in intensity and spread over many pixels, exacerbating this problem. Using the median of pixel values contaminated by such faint stars will tend to significantly overestimate the true background sky brightness in relatively dark and unpolluted areas of the sky.

To investigate the size of the contamination produced by fainter stars, we summed the flux of all objects fainter than 10th magnitude contained in contiguous 230″ × 230″ areas in the USNO-B catalog (Monet et al. 2003) in both the *B* and *R* bandpasses in a 40° × 40° area of sky centered on the north Galactic pole. (We estimate that stars of 10th magnitude and brighter are sufficiently brighter than background pixels that the median filter will generally exclude their contribution.) Although there are no *V*-band measurements included in this catalog, we assume that the *B* and *R* measurements are likely to bracket measurements in the *V* band. The results are summarized in Table 9.

Since 81%–90% of the pixels exhibit contamination levels ≤10% of the natural sky level, we infer that our per-pixel sky brightness overestimate due to unresolved background stars will in the best case be a few percent, and in the worst case be <10% in regions far from the Galactic plane. Measurements based on the 1° areas used in the 5069 point analysis described in § 2.5 will in general be less likely to suffer even this much contamination. In areas at lower Galactic latitude, where most or all pixels will be significantly contaminated, our sky brightness images clearly show contamination due to unresolved stars. Correction of this effect, where the unresolved stellar flux can exceed the natural dark night-sky flux by large factors at dark sites, would require methods that we are presently investigating and plan to describe in a later paper. However, we note that these unresolved stars are an inseparable contribution to the visual appearance of the brightness of the night sky, and for many purposes the removal of their contribution is not necessary or desirable. For the present, we leave our data uncorrected for this effect and rely where possible on visual recognition of the contaminated areas to prevent their being used to characterize light pollution measurements.

Subtracting unresolved stars from the sky background brightness measurements would lead to a significant improvement in accuracy of the all-sky brightness index, particularly when the summer Milky Way is above the horizon. When the brighter portions of the Milky Way are located within light domes, measurements of the total light from these cities will be biased, unless a compensating technique is developed.

#### 4.1.4.2. Zodiacal Light

Zodiacal light will also bias all sky brightness measurements within 1–2 hr of the end or beginning of astronomical twilight, particularly when the ecliptic is aligned at steep angles to the horizon. The gegenschein was observed on several data sets taken near local midnight. However, its contribution and that of the zodiacal band at large angles from the Sun appear to provide a negligible contribution to all overall sky brightness measurements.

#### 4.1.4.3. Variations in Airglow

Natural airglow is a very significant component of sky brightness at dark sites (Roach & Gordon 1973). Emissions resulting from oxygen, sodium, and hydroxyl excitation are common, but the [O I] line at 557.7 nm has the greatest effect on our measurements, while OH emissions in the red and infrared are effectively filtered out (see Fig. 14). Variations in airglow have a large sporadic component, apart from the general trends asso-ciated with the solar activity (sunspot) cycle, and will be difficult or impossible to model. While long-term trends in airglow have been shown to be correlated with the 10.7 cm solar flux (Walker 1988), short-term variations are unpredictable and have been observed to occur on a timescale of tens of minutes (Krisciunas 1997; Patat 2003). Natural airglow may also vary with time of night. Walker (1988) found that V-band sky brightness at dark sites decreased by about 0.4 mag in the 6 hr following the end of astronomical twilight. In cases where we have multiple data sets spanning the night at a single site, we do not find that in general this effect is verified. In fact, on several occasions the sky was observed to brighten as the night progressed. Krisciunas (1997) concluded that such short-term variations are random. To accurately measure light pollution, and particularly to monitor changes in light pollution over time, it will be necessary to account for these natural variations.

The all-sky nature of the data collected by this program allows for determining whether or not an observed change is because of natural variations or changes in size and intensity of light domes from surrounding cities and towns. If a change in sky brightness over time is observed in the entire sky (including the zenith), while the output of each city's light dome remains relatively constant, it can be safely assumed that the change is because of a variation in the natural airglow. Such changes may therefore be compensated for on a data set by data set basis when attempting to extract measurements of light pollution only (see § 3.3). At moderately and heavily light-polluted locations, light pollution levels will overwhelm more subtle changes in natural airglow, and changes in the artificial component over time will be easier to detect.

### 5. SUMMARY AND CONCLUSIONS

Precise panoramic sky brightness measurements can be produced with relatively inexpensive off-the-shelf commercial equipment. We have developed and evaluated a variety of portable automated systems costing between $5000 and $15,000 at the time of this writing, based on commercially available thermoelectrically cooled CCD cameras and robotic telescope mounts that produce measurements over the entire sky in a bandpass closely approximating the Johnson V band. Each camera system, including batteries sufficient to operate the system for several hours, is light enough that one or two workers can carry the system reasonable distances on foot. A complete observation of the entire sky requires between 22 and 35 minutes. The data are calibrated with stellar images extracted from the same images as those used for the sky brightness measurements, and they indicate a precision of 4% or better, neglecting the peculiar spectral issues pertaining to measurements of sky glow.

These systems are being used to amass a database currently including over 300 all-sky observations at over 80 locations throughout the US, primarily in lands administered by the National Park Service. The data allow complete characterization of the natural and artificial sky glow conditions from horizon to horizon, an immense improvement over the single- or few-point measurements typically reported at astronomical observatory sites by other workers using single-channel photoelectric photometers or CCD cameras mounted to large telescopes with narrow fields. Although each data set contains millions of accurate measurements, a more manageable data set consisting of measurements made on a 2°-on-center grid (producing a total of 5069 measurements over the entire sky) suffices for most purposes.

We discuss the issues arising from the varying spectral characteristics of natural and artificial components of sky glow, and their effects on the measurements produced by our system, as well as the differences between the measurements made in our system and the impression produced by the eye. We show that due to the Purkinje shift in spectral response of the dark-adapted (scotopic) eye (see Kinney 1958), yellow light sources such as high-pressure sodium produce a substantially lower amount of sky glow, on a lumen-for-lumen basis, than blue-rich light sources such as metal halide.

## 5.1. Recommended Equipment

Of the two detectors employed, the KAF 1001E appears to be superior for this effort. Compared to the KAF 261E, it possesses greater sensitivity (obviating the necessity for binning), improved linear response at low light levels, and a larger overall size requiring fewer images per data set, although it is considerably more challenging to produce a flat-field calibration for this wider field instrument. The FLI IMG 1001E electronics appear to produce an almost perfectly linear response from this larger chip, while the SBIG STL camera produced some low-level nonlinearity, even though the same detector is used in both. The SBIG camera has the advantage of accepting unregulated 12 V DC input power, requiring fewer accessories in the field and providing greater electrical efficiency. For both cameras, we used an inexpensive Nikon 50 mm f/1.8 lens set at f/2. However, the added cost of the large-format system (roughly 3 times the cost of the small-format system) may be a deterrent to investigators on a limited budget. The NexStar 5i robotic mount has functioned well in this study and is relatively inexpensive. Any notebook computer capable of running the necessary software will suffice. However, computers that can accommodate an auxiliary battery and permit the monitor to be switched off allow data collection of up to 6 hr without an external power supply.

## 5.2. Sky Quality Measurements and National Parks

The goal of measuring sky brightness for the National Park Service is to describe the quality of the nightscape, quantify how much it deviates from natural conditions, and record how it changes with time due to changes in natural conditions as well as artificial lighting in areas within and outside of the parks. This is important to the protection of the resources and values for which the parks were created, and to the minimizing of human-caused disturbances to the ecosystem. The results presented herein show that light pollution from cities can be detected at distances of over 170 km, and at locations near major sources of light pollution, very significant degradation in sky quality has been measured. The National Park Service Organic Act of 1916 (16 U.S.C., chaps. 1, 2, 3, and 4), which created the National Park Service, states as part of its mission "…to conserve the scenery and the natural and historic objects and the wildlife therein and to provide for the enjoyment of the same in such manner and by such means as will leave them unimpaired for the enjoyment of future generations." The word "unimpaired" in this mandate has in recent years given rise to a need for quantitative guidelines for what constitutes "impairment" to many natural and cultural resources.

The metrics proposed in this paper are a first attempt at quantitative descriptors that can be directly related to both visitor experience and ecosystem function. They rely on the standard methods of astronomical photometry and its instrumentation. Other measures may be developed, however, such as measurements of the illuminance of the land at various compass directions, or indices of sky illumination that include duration and timing of events above a certain threshold. These can be more directly related to ecological processes, such as predator/prey relationships in nocturnal animals, the phenology of night-blooming plants, or the nocturnal behavior of birds and other migrating animals. Further research on the effects of elevated sky brightness will be necessary to develop threshold values to determine what constitutes impairment within parks. Details on the Night Sky program of the National Park Service can be found online.[1]

The authors wish to thank G. W. Lockwood (Lowell Observatory) for many discussions and for providing resampled sky and lamp spectra; S. Levine (USNO) for assistance with the faint-star contamination analysis using the USNO-B2.0 catalog; F. H. Harris for assistance concerning CCD response characteristics; F. J. Vrba and H. C. Harris for discussions concerning the relation between stellar and surface brightness photometry; Paul Barret for helpful suggestions on the manuscript; Rebecca Stark, Daniel Elsbrock, and Cindy Duriscoe for help in field data collection; Don Davis and Steve Howell for consultation on CCD camera methods; and Chris Shaver for support of the preservation of dark night skies in the National Park Service. This research was funded in part by the US Department of Interior National Park Service.

---

[1] See http://www2.nature.nps.gov/air/lightscapes.